\documentclass[11pt]{article}
\usepackage{graphicx}
\usepackage[left=3cm,right=3cm,top=3cm,bottom=2.5cm]{geometry}

\begin{document}

\title{Occupation time of a randomly accelerated particle on the positive half axis: Results for the first five moments}
\author{Theodore  W. Burkhardt\\ Department of Physics, Temple University\\
Philadelphia, PA 19122, USA}

\maketitle


\begin{abstract}
In the random acceleration process a point particle moving in one dimension is accelerated by Gaussian white noise with zero mean. Although several fundamental statistical properties of the motion have been analyzed in detail, the statistics of occupation times is still not well understood. We consider the occupation or residence time $T_+$  on the positive $x$ axis of a particle which is randomly accelerated on the unbounded $x$ axis for a time $t$. The first two moments of $T_+$ were recently derived by Ouandji Boutcheng et al. \cite{OB}. With an alternate approach utilizing basis functions which have proved useful in other studies of randomly accelerated motion, results for the first five moments are obtained in this paper.\\\\

\end{abstract}

\maketitle

\newpage

\section{Introduction}\label{intro}
In the random acceleration process the variables, $x$ and $v$, evolve according to 
\begin{equation}
{dx\over dt}=v\,,\qquad {dv\over dt}=\eta(t)\,,\label{eqmo}
\end{equation}
where $\eta(t)$ is Gaussian white noise, with $\langle\eta(t)\rangle=0$ and $\langle\eta(t)\eta(t')\rangle=2\delta(t-t')$.
The quantities $x$ and $v$ may be interpreted as the position and velocity of a particle moving on the $x$ axis and subject to a random force. Another interpretation is to regard $v(t)$ as the position of a particle making a one-dimensional random walk and $x(t)$ as the area under the corresponding Brownian curve. Still other interpretations are found in physical applications to semiflexible polymers \cite{TWB93,TWB97,BB1,YBG}, interface growth \cite{MB1,GB,DT}, and the Burgers equation \cite{PV}. 

Several of the fundamental statistical properties of randomly accelerated motion have been anayzed in detail. The probability density $G(x,v;x_0,v_0;t)$ for propagation from initial values $(x_0,v_0)$ to $(x,v)$ in a time $t$ is known both for propagation on the unbounded line $-\infty<x<\infty$ and for propagation on the half line $x>0$ with an absorbing boundary condition at the origin \cite{TWB93,McK,MW,Sinai,Lachal1,Lachal2}. Partially absorbing and inelastic boundaries have also been considered \cite{DGL,TWB00}. In addition, there are some exact results for motion on the finite line $0<x<L\;$ \cite{FR,MP,BB2}. For recent reviews of random acceleration statistics, with emphasis on first-passage properties, see \cite{BMS,TWB14}. 

Despite this progress, the statistics of occupation times for randomly accelerated motion is still not well understood. For early work on the statistics of occupation times and applications to various stochastic processes, see  \cite{BMS,Levy,Kac,Lamperti,Cox,Godreche} and references therein. The particular occupation or residence time  considered in this paper is the time $T_+$ spent on the positive $x$ axis by a particle with initial position $x_0$ and initial velocity $v_0$, which is randomly accelerated on the unbounded $x$ axis for a time $t$. In the case of random walk statistics, both $T_+$ and $T_m$, the time at which the walker reaches its maximum excursion from the starting point, are distributed according to L\'evy's arcsine law \cite{Levy,Feller}. For random acceleration statistics Majumdar et al. \cite{MRZ} have derived the distribution of $T_m$. The distribution of $T_+$  is still unknown, but its first two moments were recently calculated by Ouandji Boutcheng et al. \cite{OB}, who obtained
\begin{eqnarray}
&&\langle T_+\rangle(x_0,v_0,t)={1\over 2}\thinspace t+{1\over 2}\thinspace\int_0^t dt'\thinspace{\rm erf}\left[{\sqrt{3}\over 2}\thinspace{x_0+v_0(t-t')\over \left(t-t'\right)^{3/2}}\right],\label{OB1}\\
&&\langle T_+\rangle(0,0,t)={1\over 2}\thinspace t\,,\label{OB2}\\
&&\langle T_+^2\rangle(0,0,t)={3^{3/2}\over 4\pi}\thinspace t^2 \approx 0.413\thinspace 497\thinspace t^2\,,\label{OB3}
\end{eqnarray}
where ${\rm erf}$ denotes the error function \cite{abramowitz}.
From the expression for $\langle T^2_+\rangle(0,0,t)$ in (\ref{OB3}) and the result $\langle T^2_m\rangle(0,0,t)={5\over 12}\thinspace t^2\approx0.416\thinspace 667\thinspace t^2$ of \cite{MRZ}, and from Monte Carlo calculations of the distributions of $T_+$ and $T_m$ reported in \cite{OB}, it is clear that the distributions of $T_+$ and $T_m$ are very similar, but not identical.  

Section \ref{genfunc} of this paper begins with the same generating-function formalism for the distribution of $T_+$ and its moments as in \cite{OB}. The generating function is then expanded in a set of basis functions which have proved useful in studying random acceleration on the half line \cite{TWB93,MW,MRZ}. To calculate the generating function, one needs to determine the expansion coefficients. In Section \ref{integeqs} this crucial step is reduced to the problem of solving an integral equation. In Section \ref{Q4thorder} the integral equation is solved to low orders in a perturbation expansion, and in Section \ref{moments} results for the first five moments of  $T_+$ are presented. Section \ref{conclusion} contains concluding remarks.

\section{Generating function $Q_p(x_0,v_0,t)$}\label{genfunc}
A quantity of central importance in our study is the generating function
\begin{equation}
Q_p(x_0,v_0,t)=\left\langle\exp\left[-p\int_0^t dt'\thinspace\theta\left(x(t')\right)\right]\right\rangle,\label{Qdef}
\end{equation}
also considered in \cite{OB}.
Here $\theta(x)$ is the usual unit step function, and the angular brackets denote an average over the possible paths of a randomly accelerated particle with initial position $x_0$ and initial velocity $v_0$. The moments of $T_+$ are related by
\begin{equation}
\langle T^n_+ \rangle(x_0,v_0,t) =  \left.(-1)^n \frac{\partial^n}{\partial p^n}Q_p(x_0,v_0,t) \right\vert_{p=0}\label{T+nfromQ}\label{derivativesofQ}
\end{equation}
to derivatives of the generating function, 
and the distribution of $T_+$, 
\begin{equation}
{\cal P}(T_+\vert x_0,v_0,t)\equiv\left\langle\delta\left(T_+ -\int_0^t dt'\thinspace\theta\left(x(t')\right)\right)\right\rangle={\cal L}^{-1}_{p\to T_+}Q_p(x_0,v_0,t)
\,,\label{distT+}
\end{equation}
is the inverse Laplace transform of the generating function.

Below, we will make use of the symmetric property
\begin{equation}
Q_p(x_0,v_0,t)=e^{-pt}Q_{-p}(-x_0,-v_0,t)\,,\label{sympropQ}
\end{equation}
which follows from the facts that (i) the residence times $T_+$ and $T_-$ on the positive and negative $x$ axes satisfy $T_++T_-=t$, and (ii) the distribution of $T_-$  for a particle with initial conditions $x_0,v_0$  is the same as the distribution of $T_+$ for a particle with initial conditions  $-x_0,-v_0$. 

The generating function defined by (\ref{Qdef}) satisfies the Fokker-Planck type differential equation \cite{OB,TWB93,MW,Kac,risken}   
\begin{equation}
\left({\partial\over\partial t}-v{\partial\over\partial x}-{\partial^2\over\partial v^2}+p\theta(x)\right)Q_p(x,v,t)=0\,,\label{Qdiffeq}
\end{equation}
as follows, for example, from path integral considerations.
To simplify the notation, the subscripts of $x_0$ and $v_0$ have been dropped in (\ref{Qdiffeq}).
Forming the Laplace transform $t\to s$ of (\ref{Qdiffeq}) and using  the initial condition 
$Q_p(x,v,0)=1$ leads to
\begin{equation}
\left(s+p\theta(x)-v{\partial\over\partial x}-{\partial^2\over\partial v^2}\right)\tilde{Q}_p(x,v,s)=1\,,\label{Qtildediffeq}
\end{equation}
where $\tilde{Q}_p(x,v,s)=\int_0^\infty dt\thinspace e^{-st}Q_p(x,v,t)$.
The Laplace transform of the symmetry property (\ref{sympropQ}) has the form
\begin{equation}
\tilde{Q}_p(x_0,v_0,s)=\tilde{Q}_{-p}(-x_0,-v_0,s+p)\,.\label{sympropQtilde}
\end{equation} 

Let us look for a solution to the differential equation  (\ref{Qtildediffeq}) in terms of the basis functions
 \begin{equation}
\psi_{s,F}(\pm v)=F^{-1/6}{\rm Ai}\left(\pm F^{1/3}v+F^{-2/3}s\right),\label{psidef}
 \end{equation}
where ${\rm Ai}(z)$ denotes the standard Airy function \cite{abramowitz}. The combined set of functions $\psi_{s,F}(v)$, $\psi_{s,F}(-v)$, with $0<F<\infty$,  is complete on the interval $-\infty< v<\infty$, and this basis has been used to good advantage in studying randomly accelerated motion on the half line \cite{TWB93,MW,MRZ}. The functions (\ref{psidef}) satisfy the Airy differential equation 
\begin{equation}
\left(s\pm F v-{\partial^2\over \partial v^2}\right)\psi_{s,F}(\pm v)=0\,,\label{diffeqpsi}
\end{equation}
the orthonormality conditions
\begin{eqnarray}
&&\int_{-\infty}^\infty dv\thinspace v\thinspace\psi_{s,F}(-v)\psi_{s,G}(-v)=\delta(F-G)\,,\label{orthonorm1}\\
&& \int_{-\infty}^\infty dv\thinspace v\thinspace\psi_{s,F}(-v)\psi_{s,G}(v)=0\,,\label{orthonorm2}
\end{eqnarray}
and the closure relation
\begin{equation}
\int_0^\infty dF\thinspace\left[\psi_{s,F}(-v)\psi_{s,F}(-v')-\psi_{s,F}(v)\psi_{s,F}(v')\right]=v^{-1}\delta(v-v')\,.\label{closure}
\end{equation}

In terms of these basis functions the most general solution of (\ref{Qtildediffeq}) which does not diverge for $x\to\pm\infty$ is given by
\begin{equation}
 \tilde{Q}_p(x,v,s)=\left\{\begin{array}{l}\displaystyle\ {1\over s}+\int_0^\infty dF\thinspace a(F)e^{Fx}\psi_{s,F}(-v)\,,\quad x<0\,,\\[0.2cm]
 \displaystyle{1\over s+p}+\int_0^\infty dF\thinspace b(F)e^{-Fx}\psi_{s+p,F}(v)\,,\quad x>0\,.\end{array}\right.\label{gensolQtilde}
 \end{equation}
 To avoid a term proportional to $\delta(x)$ when the general solution (\ref{gensolQtilde}) is substituted into differential equation (\ref{Qtildediffeq}), $ \tilde{Q}_p(x,v,s)$ must be continuous at $x=0$. This leads to the matching condition
 \begin{equation}
 {p\over s(s+p)}=\int_0^\infty dF\thinspace\left[-a(F)\thinspace\psi_{s,F}(-v)+b(F)\thinspace\psi_{s+p,F}(v)\right].\label{matchcond}
 \end{equation}
 
 \section{Integral equations for $a(F)$and $b(F)$}\label{integeqs}
To calculate the generating function $Q(x,v,t)$ using (\ref{gensolQtilde}), one must first determine the unknown functions $a(F)$ and $b(F)$. In this section separate integral equations for $a(F)$ and $b(F)$ are derived from the matching condition (\ref{matchcond}). 

With the help of the expansion
\begin{equation} 
{1\over s}= \int_0^\infty dF\thinspace F^{-3/2}\left[\psi_{s,F}(-v)+\psi_{s,F}(v)\right]\label{identity1}
\end{equation} 
in terms of the basis functions (\ref{psidef}), which is derived in Appendix A, equation (\ref{matchcond}) may be rewritten as
\begin{eqnarray}
&&{p\over s+p}\thinspace\int_0^\infty dF\thinspace F^{-3/2}\left[\psi_{s,F}(-v)+\psi_{s,F}(v)\right]\nonumber\\
&&={p\over s}\thinspace\int_0^\infty dF\thinspace F^{-3/2}\left[\psi_{s+p,F}(-v)+\psi_{s+p,F}(v)\right]\nonumber\\
&&=\int_0^\infty dF\thinspace\left[-a(F)\thinspace\psi_{s,F}(-v)+b(F)\thinspace\psi_{s+p,F}(v)\right].\label{matchcond2}\end{eqnarray}
Solving (\ref{matchcond2}) for $a$ in terms of $b$ and $b$ in terms of $a$, using the orthonormality properties (\ref{orthonorm1}),   (\ref{orthonorm2}), we obtain the coupled integral equations  
\begin{eqnarray}
 &&a(F)=-{p\over s+p}\thinspace F^{-3/2}+\int_0^\infty dG\thinspace k(F,G)\thinspace b(G)\,,\label{a(b)}\\
 &&b(F)={p\over s}\thinspace F^{-3/2}-\int_0^\infty dG\thinspace k(G,F)\thinspace a(G)\,,\label{b(a)}
\end{eqnarray}
where
\begin{eqnarray}
k(F,G)&\equiv&\int_{-\infty}^\infty dv\thinspace v\thinspace\psi_{s,F}(-v)\psi_{s+p,G}(v)\nonumber\\
&=&-p(FG)^{-1/6}(F+G)^{-4/3}{\rm Ai}\left({(s+p)F+sG\over (F+G)^{1/3}(FG)^{2/3}}\right).\label{kdef}
\end{eqnarray}
A derivation of the result (\ref{kdef}) is given in Appendix B. To decouple (\ref{a(b)}) and  (\ref{b(a)}), we substitute (\ref{b(a)}) on the right side of (\ref{a(b)}). This leads to an inhomogeneous Fredholm integral equation of the second kind, with a symmetric kernel, 
\begin{eqnarray}
&&a(F)=a_0(F)+\int_0^\infty dG\thinspace K(F,G)\thinspace a(G)\,,\label{a(a)}\label{integeqfora}\\
&&a_0(F)=-{p\over s+p}\thinspace F^{-3/2}+{p\over s}\thinspace\int_0^\infty dG\thinspace k(F,G)G^{-3/2}\,,\label{a0def}\\
&& K(F,G)=-\int_0^\infty dH\thinspace k(F,H)\thinspace k(G,H)\,,\label{Kdef}
\end{eqnarray}
for the unknown function $a(F)$. The corresponding integral equation for $b(F)$ is the same as for $a(F)$, except that the variables $s$ and $s+p$ are switched. (Under this switching, the prefactor $p=(s+p)-s$ in the last line of (\ref{kdef}) transforms to $ s-(s+p)=-p$.) 

\section{Calculation of $Q_p(0,0,t)$ to 4th order in $p$}\label{Q4thorder}
Together with equations (\ref{gensolQtilde}),  (\ref{kdef}), (\ref{a0def}), and  (\ref{Kdef}), the solution to the integral equation (\ref{integeqfora}) completely determines the generating function. However, solving the integral equation exactly appears extremely difficult, due to the complicated kernel (\ref{kdef}), (\ref{Kdef}). In this section we calculate $a(F)$ and $\tilde{Q}_p(0,0,s)$ perturbatively to fourth order in powers of $p$, confirming and extending the second order results in \cite{OB}. With little extra work the fourth order results are extended to fifth order in the following section.

Expanding the quantity $a_0(F)$, defined in (\ref{kdef}) and (\ref{a0def}), to fourth order in $p$, one finds
\begin{eqnarray}
&&a_0(F)=-{p\over s}\left(1-{p\over s}+{p^2\over s^2}-{p^3\over s^3} \right)F^{-3/2}\nonumber\\
&&\quad -{p^2\over s}\int_0^\infty dG\thinspace G^{-3/2}(FG)^{-1/6}(F+G)^{-4/3}\nonumber\\ && \quad\times\left[
{\rm Ai}\left(s\left(F^{-1}+G^{-1}\right)^{2/3}\right)\right.\nonumber\\
&&\quad +p\thinspace G^{-1}\left(F^{-1}+G^{-1}\right)^{-1/3}{\rm Ai'}\left(s\left(F^{-1}+G^{-1}\right)^{2/3}\right)\nonumber\\
&&\quad\left. +{1\over 2}\thinspace p^2 G^{-2}\left(F^{-1}+G^{-1}\right)^{-2/3}{\rm Ai''}\left(s\left(F^{-1}+G^{-1}\right)^{2/3}\right)\right]+{\rm O}\left(p^5\right).\nonumber\\ \label{a0def1}
\end{eqnarray}
On changing the integration variable from $G$ to $z=1+F/G$, integrating the term involving ${\rm Ai'(\xi)}$ by parts, and utilizing the Airy differential equation ${\rm Ai''}(\xi)=\xi{\rm Ai}(\xi)$, equation (\ref{a0def1}) takes the form
\begin{eqnarray}
&&a_0(F)=-{p\over s}\left(1-{p\over s}+{p^2\over s^2}-{p^3\over s^3} \right)F^{-3/2}\nonumber\\
&&\quad -{p^2\over s}F^{-13/6}\int_1^\infty dz\thinspace{\rm Ai}\left(sF^{-2/3}z^{2/3}\right)\nonumber\Bigg\{(z-1)z^{-4/3}\nonumber\\
&&\quad -{3p\over 2s}{d\over dz}\left[(z-1)^2z^{-4/3}\right]+{1\over2}p^2sF^{-2}(z-1)^3z^{-4/3}\Bigg\}
+{\rm O}\left(p^5\right).\label{a0def2}
\end{eqnarray}
With the help of {\it Mathematica}, all of the integrals in (\ref{a0def2}) can be evaluated explictly in terms of generalized hypergeometric functions.

The integral equation $a=a_0+\int Ka$ in (\ref{integeqfora}) has the iterative solution $a=a_0+\int Ka_0 +\int\int KKa_0+ ...\;$ Since $a_0(F)$ in equations (\ref{kdef})-(\ref{Kdef}) is O($p$) and $K(F,G)$ is of order O$\left(p^2\right)$,  we only need to consider the first two terms, $a_0+\int Ka_0$, to obtain $a(F)$ to 4th order in $p$. The first term, $a_0(F)$, is shown to 4th order in $p$ in (\ref{a0def2}). The second term, $\int Ka_0$, has the expansion 
\begin{equation}
\int Ka_0\equiv\int_0^\infty dG\thinspace K(F,G)\thinspace a_0(G)= a_1(F)+a_2(F)+a_3(F)+{\rm O}\left(p^5\right),\label{a1+a2}
\end{equation}
\begin{eqnarray}
 &&a_1(F)={p^3\over s}\left(1-{p\over s}\right)\nonumber\\ 
&&\qquad\times \int_0^\infty dG\int_0^\infty dH\thinspace G^{-3/2}(FH^2G)^{-1/6}(F+H)^{-4/3}(G+H)^{-4/3}\nonumber\\
&&\qquad\times{\rm Ai}\left(s\left(F^{-1}+H^{-1}\right)^{2/3}\right){\rm Ai}\left(s\left(G^{-1}+H^{-1}\right)^{2/3}\right),\label{a1def}\\
&&a_2(F)={p^4\over s}\int_0^\infty dG\int_0^\infty dH\thinspace
G^{-3/2}(FH^2G)^{-1/6}(F+H)^{-4/3}(G+H)^{-4/3}\nonumber\\
&&\quad\times H^{-1}\left[\left(F^{-1}+H^{-1}\right)^{-1/3}{\rm Ai'}\left(s\left(F^{-1}+H^{-1}\right)^{2/3}\right){\rm Ai}\left(s\left(G^{-1}+H^{-1}\right)^{2/3}\right)\right.\nonumber\\
&&\quad\left. +\left(G^{-1}+H^{-1}\right)^{-1/3}{\rm Ai}\left(s\left(F^{-1}+H^{-1}\right)^{2/3}\right){\rm Ai'}\left(s\left(G^{-1}+H^{-1}\right)^{2/3}\right)\right],\nonumber\label{a2def}\\
&&a_3(F)={p^4\over s}\int_0^\infty dG\int_0^\infty dH
(FH^2G)^{-1/6}(F+H)^{-4/3}(G+H)^{-4/3}\nonumber\\
&&\qquad\qquad\times\thinspace {\rm Ai}\left(s\left(F^{-1}+H^{-1}\right)^{2/3}\right){\rm Ai}\left(s\left(G^{-1}+H^{-1}\right)^{2/3}\right)\nonumber\\
&&\qquad\qquad\times
G^{-13/6}\thinspace\int_1^\infty dz\thinspace(z-1)z^{-4/3}{\rm Ai}\left(sG^{-2/3}z^{2/3}\right).\label{a3def}
\end{eqnarray}

In analyzing the integrals in (\ref{a1def})-(\ref{a3def}), we make the substitutions $H=F/h$ and $G=H/g=F/(gh)$, which lead to 
\begin{eqnarray}
&&a_1(F)={p^3\over s}\left(1-{p\over s}\right)F^{-17/6}\int_0^\infty dg\int_0^\infty dh\thinspace g\thinspace h^{5/3}(1+g)^{-4/3}(1+h)^{-4/3}\nonumber\\
&&\quad\times\thinspace{\rm Ai}\left(sF^{-2/3}(1+h)^{2/3}\right){\rm Ai}\left(sF^{-2/3}h^{2/3}(1+g)^{2/3}\right),\label{a1def2}\\
&&a_2(F)={p^4\over s}F^{-7/2}\int_0^\infty dg\int_0^\infty dh\thinspace g\thinspace h^{5/3}(1+g)^{-4/3}(1+h)^{-4/3}\nonumber\\
&&\quad\times\left[ h (1+h)^{-1/3}{\rm Ai'}\left(sF^{-2/3}(1+h)^{2/3}\right){\rm Ai}\left(sF^{-2/3}h^{2/3}(1+g)^{2/3}\right)\right. 
\nonumber\\ &&\quad\left. +h^{2/3} (1+g)^{-1/3}{\rm Ai}\left(sF^{-2/3}(1+h)^{2/3}\right){\rm Ai'}\left(sF^{-2/3}h^{2/3}(1+g)^{2/3}\right)\right] \,,\nonumber \label{a2def2}\\
&&a_3(F)={p^4\over s}\thinspace F^{-7/2}\int_0^\infty dg\int_0^\infty dh\thinspace g^{5/3}\thinspace h^{7/3}(1+g)^{-4/3}(1+h)^{-4/3}\nonumber\\
&&\quad\times\thinspace{\rm Ai}\left(sF^{-2/3}(1+h)^{2/3}\right){\rm Ai}\left(sF^{-2/3}h^{2/3}(1+g)^{2/3}\right) \nonumber\\
&&\quad\times\thinspace \int_1^\infty dz\thinspace(z-1)z^{-4/3}{\rm Ai}\left(sF^{-2/3}(hg)^{2/3}z^{2/3}\right). \label{a3def2}
 \end{eqnarray}
 
 We now specialize to the case of a particle which begins at the origin with velocity zero. According to equations (\ref{psidef}) and (\ref{gensolQtilde}), the generating function $\tilde{Q}_p\left(x_0,v_0,s\right)$ for these initial conditions is given by
 \begin{eqnarray}
&& \tilde{Q}_p(0,0,s)-{1\over s}=\int_0^\infty dF\thinspace a(F)\psi_{s,F}(0)\nonumber\\&&\qquad =
 \int_0^\infty dF\thinspace a(F)\thinspace F^{-1/6} {\rm Ai}\left(sF^{-2/3}\right)\nonumber\\
 &&\qquad ={3\over 2}\ s^{5/4}\int_0^\infty dx\thinspace a\left(\left({s\over x}\right)^{3/2}\right)\thinspace x^{-9/4}\thinspace{\rm Ai}(x)\,,\;\; x=sF^{-2/3}\,.\label{Qfroma}
\end{eqnarray}

Let us denote the contributions to the right side of (\ref{Qfroma}) from $a_0(F)$,..., $a_3(F)$  in equations (\ref{a0def2}), (\ref{a1def2}), (\ref{a2def2}), and (\ref{a3def2}) by $\tilde{Q}^{(0)}_p(0,0,s)$,..., $\tilde{Q}^{(3)}_p(0,0,s)$.
Substituting expression (\ref{a0def2}) for $a_0(F)$ in (\ref{Qfroma})  and making use of the integrals
\begin{eqnarray}
&& \int_0^\infty dx\thinspace{\rm Ai}(x)={1\over 3}\,,\label{integ1}\\
&& \int_0^\infty dx\thinspace x{\rm Ai}(x){\rm Ai}(ax)={1\over 2\pi\sqrt{3}}\thinspace{a-1\over a^3-1}\,,\label{integ2}\\
&&\int_0^\infty dx\thinspace x^4{\rm Ai}(x){\rm Ai}(ax)={\sqrt{3}\over\pi}\thinspace(a+1)\left({a-1\over a^3-1}\right)^3\,,\label{integ3}
 \end{eqnarray}
one obtains
 \begin{eqnarray}
 &&\tilde{Q}^{(0)}_p(0,0,s)=-{p\over 2s^2}+{3^{3/2}\over 4\pi}\thinspace{p^2\over s^3}
 +\left({3^{5/2}\over 8\pi}-1\right){p^3\over s^4}\nonumber\\&&\qquad + \left({3^{7/2}\over 8\pi}-{3\over 2}\right){p^4\over s^5}
 +{\rm O}\left(p^5\right).\label{Q(0)} 
\end{eqnarray}

Substituting expression (\ref{a1def2}) for $a_1(F)$ in (\ref{Qfroma}) leads to the triple integral
\begin{eqnarray}
 &&\tilde{Q}^{(1)}_p(0,0,s)={3\over 2}{p^3\over s^4}\left(1-{p\over s}\right)\int_0^\infty dx\thinspace x^2{\rm Ai}(x)\nonumber\\
  &&\quad\times\thinspace\int_0^\infty dg\int_0^\infty dh\thinspace g\thinspace h^{5/3}(1+g)^{-4/3}(1+h)^{-4/3}\nonumber\\
 &&\quad\times\thinspace{\rm Ai}\left(x(1+h)^{2/3}\right){\rm Ai}\left(xh^{2/3}(1+g)^{2/3}\right). \label{Q(1)}
\end{eqnarray}
Integrating over $g$ analytically and then over $h$ and $x$ numerically, using {\it Mathematica}, gave a result consistent with 
\begin{eqnarray}
 &&\tilde{Q}^{(1)}_p(0,0,s)=\left({5\over 4}-{3^{5/2}\over 4\pi}\right){p^3\over s^4}\left(1-{p\over s}\right) \label{Q(1)2}
\end{eqnarray}
with an uncertainty in the numerical prefactor less than $10^{-12}$. That (\ref{Q(1)2}) is not just an excellent approximation but is exact is shown in the next section. 

Substituting expression (\ref{a2def2}) for $a_2(F)$ in (\ref{Qfroma}) leads to the triple integral
\begin{eqnarray}
 &&\tilde{Q}^{(2)}_p(0,0,s)={3\over 2}{p^4\over s^5}\int_0^\infty dx\thinspace x^3{\rm Ai}(x)\nonumber\\
 &&\quad\times\thinspace\int_0^\infty dg\int_0^\infty dh\thinspace g\thinspace h^{5/3}(1+g)^{-4/3}(1+h)^{-4/3}\nonumber\\
 &&\quad\times\left[ h (1+h)^{-1/3}{\rm Ai'}\left(x(1+h)^{2/3}\right){\rm Ai}\left(xh^{2/3}(1+g)^{2/3}\right)\right. 
\nonumber\\ &&\quad\left. +h^{2/3} (1+g)^{-1/3}{\rm Ai}\left(x(1+h)^{2/3}\right){\rm Ai'}\left(xh^{2/3}(1+g)^{2/3}\right)\right].\label{Q(2)}
\end{eqnarray}
This integral can be written more compactly, without derivatives of Airy functions, by rewriting the square bracket as 
\begin{equation}
\left[\; ...\;\right]=\textstyle{3\over 2}\displaystyle x^{-1}\left(h{\partial\over\partial h}-
 g{\partial\over\partial g}\right){\rm Ai}\left(x(1+h)^{2/3}\right){\rm Ai}\left(xh^{2/3}(1+g)^{2/3}\right)\label{rewrite}
 \end{equation}
 and then integrating by parts.
 This yields 
\begin{eqnarray}
&&\tilde{Q}^{(2)}_p(0,0,s)=-{3\over 2}{p^4\over s^5}\int_0^\infty dx\thinspace x^2{\rm Ai}(x)\nonumber\\
&&\;\;\times\thinspace\int_0^\infty dg\int_0^\infty dh\thinspace g\thinspace h^{5/3}(1+g)^{-4/3}(1+h)^{-4/3}\nonumber\\
&&\;\;\times\left(1+{2\over 1+h}-{2\over 1+g}\right){\rm Ai}\left(x(1+h)^{2/3}\right){\rm Ai}\left(xh^{2/3}(1+g)^{2/3}\right).\label{Q(2)2alternate}
\end{eqnarray}

Performing the integral in (\ref{Q(2)2alternate}) over $g$ analytically and then over $h$ and $x$ numerically, using {\it Mathematica}, one finds that the contributions of the terms $2/(1+h)$ and $-2/(1+g)$ in the integrand cancel on integration with an uncertainty less than $10^{-14}$. Presumably the cancellation is exact, although we have not succeeded in showing  this analytically. Thus, within this minute uncertainty, $\tilde{Q}^{(2)}_p(0,0,s)$ is entirely determined by the integral of the first of the sum of three terms.
Since the integral of that term is the same as in (\ref{Q(1)}) and (\ref{Q(1)2}), apart from the prefactor, we conclude that
\begin{eqnarray}
 &&\tilde{Q}^{(2)}_p(0,0,s)=\left({3^{5/2}\over 4\pi}-{5\over 4}\right)\thinspace{p^4\over s^5} \,.\label{Q(2)2}
\end{eqnarray}

 Substituting expression (\ref{a3def2}) for $a_3(F)$ in (\ref{Qfroma}) leads to the quadruple integral
\begin{eqnarray}
 &&\tilde{Q}^{(3)}_p(0,0,s)={3\over 2}{p^4\over s^5}\int_0^\infty dx\thinspace x^3{\rm Ai}(x)\nonumber\\
 &&\quad\times\thinspace\int_0^\infty dg\int_0^\infty dh\thinspace g^{5/3}h^{7/3}(1+g)^{-4/3}(1+h)^{-4/3}\nonumber\\
 &&\quad\times\thinspace{\rm Ai}\left(x(1+h)^{2/3}\right){\rm Ai}\left(xh^{2/3}(1+g)^{2/3}\right)
\nonumber\\ &&\quad\times\int_1^\infty dz\thinspace(z-1)z^{-4/3}{\rm Ai}\left(x g^{2/3}h^{2/3}z^{2/3}\right).\label{Q(3)}
\end{eqnarray}
Integrating over $z$ analytically and then over $g$, $h$, and $x$ numerically, using {\it Mathematica}, gave  
\begin{equation}
\tilde{Q}^{(3)}_p(0,0,s)\equiv\epsilon\thinspace{p^4\over s^5}\,,\quad \epsilon =0.000\thinspace872\thinspace073\thinspace 2\,,\label{Q(3)2}
\end{equation} believed to be correct to the number of digits shown. This completes the evaluation of all the contributions to $\tilde{Q}_p(0,0)$ through order $p^4$.

\section{Moments of the Occupation Time $T_+$}\label{moments}

Continuing to specialize to initial conditions $x_0=0$, $v_0=0$,  we note that  the generating functions $Q_p(0,0,t)$ and $e^{pt/2}Q_p(0,0,t)$ defined by (\ref{Qdef}) have the expansions 
\begin{eqnarray}
&&Q_p(0,0,t)=\sum_{n=0}^\infty{(-p)^n\over n!}\left\langle T_+^n\right\rangle,\label{Qexpansion}\\
&&e^{pt/2}Q_p(0,0,t)=\sum_{n=0}^\infty{(-p)^n\over n!}\left\langle \left(T_+-\textstyle{1\over 2}t\right)^n\right\rangle,\label{altQexpansion}
\end{eqnarray}
in terms of moments of $T_+$. Here and throughout the remainder of the paper we use the shorthand  $\langle T_+^n\rangle\equiv\langle T_+^n\rangle(0,0,t)$.
According to the symmetry property (\ref{sympropQ}), $e^{pt/2}Q_p(0,0,t)=e^{-pt/2}Q_{-p}(0,0,t)$ is an even function of $p$. Together with (\ref{altQexpansion}), this implies
\begin{equation} 
\left\langle \left(T_+- \textstyle{1\over 2}t\right)^n\right\rangle=0\quad {\rm for}\ n\ {\rm odd}\,.\label{oddmoments}
\end{equation}
This result can be understood very simply. For a particle which begins at rest at the origin, the residence times $T_+$ and $T_-=t-T_+$ on the positive and negative axes are identically distributed.  Thus, $T_+-{1\over 2}t$ and $T_- -\textstyle{1\over 2}t=-(T_+-{1\over 2}t)$ also have identical distributions, which leads directly to (\ref{oddmoments}). 

For $n$=1, 3, 5, and 7, equation (\ref{oddmoments}) implies
\begin{eqnarray}
\left \langle T_+\right\rangle&=&\textstyle{1\over 2}\thinspace t\,,\label{T1easy}\\
\left \langle T_+^3\right\rangle&=&\textstyle{3\over 2}\thinspace\left\langle T_+^2\right\rangle t -{1\over 4}t^3 \,,\label{T3easy}\\
\left \langle T_+^5\right\rangle&=&\textstyle{5\over 2}\thinspace\left\langle T_+^4\right\rangle t-{5\over 2}\thinspace\left\langle T_+^2\right\rangle t^3+ {1\over 2}t^5\,.
\label{T5easy}
\end{eqnarray}
In general $\left \langle T_+^n\right\rangle$ for $n$ odd is determined by the $0^{\rm th}$, $2^{\rm nd}$, ..., $(n-1)^{\rm st}$ moments of  $T_+$.

Combining the contributions (\ref{Q(0)}), (\ref{Q(1)2}), (\ref{Q(2)2}), and (\ref{Q(3)2}) to $\tilde{Q}_p(0,0,s)$ through order $p^4$, performing the inverse Laplace transform $s\to t$, comparing with (\ref{derivativesofQ}) or 
(\ref{Qexpansion}) to identify the moments, and making use of (\ref{T5easy}), we obtain $Q_P(0,0,t)$ to order $p^5$ and the corresponding 5 moments of $T_+$.
In terms of the quantity 
\begin{equation}
\epsilon =0.000\thinspace872\thinspace073\thinspace 2,\label{epsilon}
\end{equation} 
defined in (\ref{Q(3)}) and (\ref{Q(3)2}) and obtained by numerical integration,
\begin{eqnarray}
&&\tilde{Q}_p(0,0,s)={1\over s} -{1\over 2}\thinspace{p\over s^2}+{3^{3/2}\over 4\pi}\thinspace{p^2\over s^3}
 -\left({3^{5/2}\over 8\pi}-{1\over 4}\right){p^3\over s^4}\nonumber\\&&\qquad\qquad + \left({7\cdot 3^{5/2}\over 8\pi}-4+\epsilon\right)\thinspace{p^4\over s^5}\nonumber\\&&\qquad\quad -\left({95\cdot 3^{3/2}\over 16\pi}-{19\over 2}+{5\over 2}\thinspace\epsilon\right)\thinspace{p^5\over s^6}
 +{\rm O}\left(p^6\right),\label{Qps}\\
 &&Q_p(0,0,t)=1 -{1\over 2}\thinspace {pt\over 1!}+{3^{3/2}\over 4\pi}\thinspace{(pt)^2\over 2!}
 -\left({3^{5/2}\over 8\pi}-{1\over 4}\right){(pt)^3\over 3!}\nonumber\\&&\qquad\qquad + \left({7\cdot 3^{5/2}\over 8\pi}-4+\epsilon\right)\thinspace{(pt)^4\over 4!}\nonumber\\&&\qquad\quad -\left({95\cdot 3^{3/2}\over 16\pi}-{19\over 2}+{5\over 2}\epsilon\right)\thinspace{(pt)^5\over 5!}+{\rm O}\left(p^6\right),
 \label{Qpt}\\
&&\left \langle T_+\right\rangle=\textstyle{1\over 2}\thinspace t\,,\label{T1}\\
&&\left\langle T_+^2\right\rangle= {3^{3/2}\over 4\pi}\thinspace t^2=0.413\thinspace496\thinspace672\thinspace t^2\,,\label{T2}\\
&& \left \langle T_+^3\right\rangle=\left({3^{5/2}\over 8\pi}-{1\over 4}\right)\thinspace t^3=0.370\thinspace245\thinspace007\thinspace t^3\,,\label{T3}\\
&& \left \langle T_+^4\right\rangle= \left({7\cdot 3^{5/2}\over 8\pi}-4+\epsilon\right)t^4=0.342\thinspace587\thinspace125\thinspace t^4\,,\label{T4}\\ &&\left \langle T_+^5\right\rangle= \left({95\cdot 3^{3/2}\over 16\pi}-{19\over 2}+{5\over 2}\thinspace\epsilon\right) t^5=0.322\thinspace726\thinspace133\thinspace t^5\,.\label{T5}
\end{eqnarray}

These are our main findings. The results (\ref{T1}) and (\ref{T2}) for the first two moments are consistent with \cite{OB}, and the
third moment (\ref{T3}) follows immediately from the second moment and (\ref{T3easy}). That the O$\left(p^3\right)$ calculation in the preceding section leads to the same result for the third moment proves our conjecture that the quantity $\tilde{Q}_p^{(1)}(0,0,s)$ defined in (\ref{Q(1)}) is exactly given by (\ref{Q(1)2}). The results (\ref{T4}) and (\ref{T5}) for the fourth and fifth moments depend on the value (\ref{epsilon}) of $\epsilon$ obtained by numerical integration and are believed to be exact to the number of digits shown. 

These results for the moments can also be summarized as 
\begin{equation}   
\left \langle\left(T_+ -\textstyle{1\over 2}t\right)^n\right\rangle=\left\{\begin{array}{l}0\,,\;\; n\;{\rm odd}, \\[0.15cm]\displaystyle\left({3^{3/2}\over 4\pi}-{1\over 4}\right)\displaystyle\thinspace t^2=0.163\thinspace 496\thinspace 672\thinspace t^2\,,\;\; n=2,\\[0.35cm]
\displaystyle\left({3^{7/2}\over 4\pi}-{59\over 16}+\epsilon\right)t^4=0.034\thinspace842\thinspace117\thinspace t^4\,,\;\; n=4
\end{array}\right.\label{momentsTminustdiv2}
\end{equation}

For comparison we show the generating function and moments of the time $T_m$ at which a particle, initially at rest at the origin and randomly accelerted for a time $t$, reaches its maximum displacement \cite{MRZ}. In terms of the hypergeometric function $_2F_1$ and the modified Bessel function $I_\nu$ \cite{abramowitz},
\begin{eqnarray}
&&\tilde{Q}^{(m)}_p(0,0,s)=s^{-1}\thinspace_2F_1(\textstyle{1\over 4},1;{1\over 2};-p/s)\,.\\
&&Q^{(m)}_p(0,0,t)=\textstyle{2^{-1/2}}\Gamma\left(\textstyle{3\over 4}\right)(pt)^{1/4}e^{-pt/2}I_{-{1\over 4}}(pt/2)\,,\\ 
&&\left \langle T_m^n\right\rangle=\textstyle{\Gamma\left({1\over 2}\right)\over\Gamma\left({1\over 2}+n\right)}
{\Gamma\left({1\over 4}+n\right)\over\Gamma\left({1\over 4}\right)}\,t^n\,,\\
&&\left \langle T_m\right\rangle=\textstyle{1\over 2}\thinspace t\,,\label{Tm1}\\
&&\left\langle T_m^2\right\rangle=\textstyle {5\over 12}\thinspace t^2=0.416\thinspace666\thinspace667\thinspace t^2\,,\label{Tm2}\\
&& \left \langle T_m^3\right\rangle=\textstyle{3\over 8}\thinspace t^3=0.375\thinspace000\thinspace000\thinspace t^3\,,\label{Tm3}\\
&& \left \langle T_m^4\right\rangle= \textstyle{39\over 112}\thinspace t^4 =0.348\thinspace 214\thinspace 286\thinspace t^4\,,\label{Tm4}\\ 
&&\left \langle T_m^5\right\rangle= \textstyle{221\over 672}\thinspace t^5=0.328\thinspace 869\thinspace048\thinspace t^5\,,\label{Tm5}\\ 
&&\left \langle\left(T_m -\textstyle{1\over 2}t\right)^n\right\rangle=\left\{\begin{array}{l}0\,,\;\; n\;{\rm odd},\\[0.1cm]
\textstyle{2^{1/2-n}}{\Gamma\left({1\over 2}\right)\Gamma\left({1\over 2}+{n\over 2}\right)\over 
\Gamma\left({1\over 4}\right)\Gamma\left({3\over 4}+{n\over 2}\right)}\,t^n\,,\;\; n\;{\rm even}\,,\\[0.35cm]
\textstyle{1\over 6}\thinspace t^2=0.166\thinspace 666\thinspace 667\thinspace t^2\,,\;\; n=2\,,\\[0.1cm]
\textstyle{1\over 28}\thinspace t^4=0.035\thinspace\thinspace714\thinspace 286\thinspace t^4\,,\;\; n=4\,.
\end{array}\right.\label{momentsTmminustdiv2}
\end{eqnarray} 
Apart from the first moment, all of the moments of $T_+$ are slightly smaller than the corresponding moments of $T_m$, in accordance with the conclusion of \cite{OB} that the distributions of $T_+$ and $T_m$ are very similar but not identical.  

\section{Concluding Remarks}\label{conclusion}
We close by comparing the approach of this paper with that of Ouandji Boutcheng et al. \cite{OB}. In both papers moments of $T_+$ are calculated by solving an integral equation to low orders in $p$, but the integral equations considered in the two papers are different. 

The results of \cite{OB} are based on the integral equation 
\begin{equation}
Q_p(x,v,t)=1-p\int_0^t dt'\int_0^\infty dx'\int_{-\infty}^\infty dv'\, Q_p(x',v',t')G(x',v';x,v;t-t')\,.\label{OBinteq}
\end{equation}
It is obtained by integrating differential equation (\ref{Qdiffeq}) with the Green's function \cite{TWB93}
\begin{eqnarray}
&&G(x',v';x,v;t)=\left(3^{1/2}/2\pi t^2\right)\nonumber\\
&&\qquad\times\exp\left[-3(x'-x-v't)(x'-x-vt)/t^3-(v'-v)^2/t\right],\label{Gdef}
\end{eqnarray}
which is the probability density for propagation on the unbounded $x$ axis from $(x,v)$ to $(x',v')$ in a time $t$. The expressions (\ref{OB1})-(\ref{OB3}) for the first two moments of $T_+$ follow from iterating (\ref{OBinteq}) to order $p^2$ and using the relation  (\ref{T+nfromQ}) between the moments and the generating function.  Due to the complicated kernel and the triple integral\footnote{The integration over $t'$  in (\ref{OBinteq}) can be eliminated by forming the Laplace transform $t\to s$ of (\ref{OBinteq}) and using the Laplace transform of $G$ given in equation (8) of \cite{TWB93}. Even with these steps it still appears extremely difficult to solve the integral equation analytically beyond order $p^2$.} on the right-hand side of (\ref{OBinteq}), obtaining analytic results beyond order $p^2$ with this approach does not seem feasible. Since the kernel in (\ref{OBinteq}) is proportional to $p$, a fourth order calculation requires four iterations of the integral equation, i.e. evaluation of a 12-fold integral.

The results of this paper, on the other hand, were derived from integral equation  (\ref{integeqfora}) for the expansion coefficient $a(F)$ in the general solution (\ref{gensolQtilde}) for the generating function.  In contrast to the triple integral in (\ref{OBinteq}), there is only a single integral on the right-hand side of (\ref{integeqfora}). However, the complicated kernel $K(F,G)$, shown in (\ref{kdef}) and (\ref{Kdef}), involves special functions and an additional integral over the variable $H$. Since $a_0(F)$ and $K(F,G)$ are of order $p$ and $p^2$, respectively, only one iteration of the integral equation (\ref{integeqfora}) is required to obtain $a(F)$ to fourth order. Even so, we needed numerical integration to evaluate the fourth-order contribution (\ref{Q(3)}).

Each of the integral equations (\ref{integeqfora}) and (\ref{OBinteq}) determines not only the first few moments of  $T_+$ but the entire distribution (\ref{distT+}) of $T_+$. The determination of this distribution is a challenging unsolved problem.

\section*{Appendix A: Derivation of the identity (\ref{identity1})}

Expression (\ref{identity1}) may be checked quickly and non-rigorously by integrating over $F$ numerically for a variety of numerical values of  $v$.
It follows analytically from integrating the closure relation (\ref{closure}) over all $v'$ to obtain 
\begin{equation}
1= v\int_0^\infty dF\thinspace F^{-1/2}\left[\psi_{s,F}(-v)-\psi_{s,F}(v)\right]\label{identity2}
\end{equation}
and rewriting this as 
\begin{equation}
1= \left(s-{\partial^2\over\partial v^2}\right)I(v)\,,\quad I(v)\equiv\int_0^\infty dF\thinspace F^{-3/2}\left[\psi_{s,F}(-v)+\psi_{s,F}(v)\right],\label{identity3}
\end{equation}
with the help of the Airy differential equation (\ref{diffeqpsi}). Differential equation (\ref{identity3}) for $I(v)$ has the solution $I(v)=s^{-1}+A\exp\left(\sqrt{s}v\right)+B\exp\left(-\sqrt{s}v\right)$, where, however, the constants $A$ and $B$ both vanish, since $I(v)$, as defined by the integral in (\ref{identity3}), remains finite for $v\to\pm\infty$. Thus, $I(v)=s^{-1}$, which, together with the definition of $I(v)$ in (\ref{identity3}), establishes (\ref{identity1}).  

 We note that (\ref{identity1}) is also consistent with the exact result
\begin{equation}
\int_0^\infty dF\thinspace F^{-3/2}\psi_{s,F}(-v)=\left\{\begin{array}{l}(2s)^{-1}\left(2-e^{-\sqrt{3s}\thinspace|v|}\right)\,,\\(2s)^{-1}e^{-\sqrt{3s}\thinspace|v|}\,, \end{array}\right.\begin{array}{l}v>0\,,\\ v<0\,.\end{array}\label{exactintegral}
\end{equation}
It may be derived by first calculating $\tilde{Q}_p(0,v,s)$ to first order in $p$, using the upper line of (\ref{gensolQtilde}) and equation (\ref{a(b)}), which imply
\begin{equation}
\tilde{Q}_p(0,v,s)={1\over s}-{p\over s}\thinspace\int_0^\infty dF\thinspace F^{-3/2}\psi_{s,F}(-v)+{\rm O}\left(p^2\right).\label{Qtildetofirstorder}
\end{equation}
Differentiating this expression with respect to $p$, as in (\ref{T+nfromQ}), we then obtain 
 \begin{equation}
\int_0^\infty dt\thinspace e^{-st}\langle T_+\rangle(0,v,t)={1\over s}\thinspace\int_0^\infty dF\thinspace F^{-3/2}\psi_{s,F}(-v)\label{LaptranT+} 
\end{equation}
for the Laplace transform of $\langle T_+\rangle(0,v,t)$. Equating this result   
to the Laplace transform of expression (\ref{OB1}) for $\langle T_+\rangle(0,v,t)$ and explicitly evaluating the latter leads directly to (\ref{exactintegral}). By combining (\ref{exactintegral}) with the Airy differential equation 
(\ref{diffeqpsi}), the integrals $\int_0^\infty dF\thinspace F^{-n}\psi_{s,F}(-v)$, where $n={1\over 2}$, ${3\over 2}$, ${5\over 2}$, ... , can all be evaluated analytically. 

\section*{Appendix B: Evaluation of the integral in equation (\ref{kdef})}\label{integral}
Combining the relation
\begin{equation}
\int_{-\infty}^\infty dv\thinspace\left[{\partial^2\over \partial v^2}\thinspace\psi_{s,F}(-v)\right]\psi_{s+p,G}(v)=\int_{-\infty}^\infty dv\thinspace\psi_{s,F}(-v){\partial^2\over \partial v^2}\thinspace\psi_{s+p,G}(v),
\end{equation}
which follows from integration by parts, with the Airy differential equation (\ref{diffeqpsi}) leads to
\begin{eqnarray}
k(F,G)&\equiv&\int_{-\infty}^\infty dv\thinspace v\thinspace\psi_{s,F}(-v)\psi_{s+p,G}(v)\nonumber\\&=&-p(F+G)^{-1}\int_{-\infty}^\infty dv\thinspace\psi_{s,F}(-v)\psi_{s+p,G}(v)\,.\label{C1}
\end{eqnarray}
With the help of definition (\ref{psidef}) and the integral representation 
\cite{abramowitz}
\begin{equation}
{\rm Ai}(z)={1\over 2\pi}\int_{-\infty}^\infty dq\thinspace e^{i\left({1\over 3}q^3+qz\right)}\,,\label{integralrep}
\end{equation}
the integral on the right side of (\ref{C1}) can be written as 
\begin{eqnarray}
&&\int_{-\infty}^\infty dv\thinspace\psi_{s,F}(-v)
\psi_{s+p,G}(v)=(2\pi)^{-2}(FG)^{-1/6}\int_{-\infty}^\infty dv\int_{-\infty}^\infty dk\int_{-\infty}^\infty d\ell\nonumber\\ &&\quad\times\exp\left\{i\left[{1\over 3}k^3+k\left(-F^{1/3}v+sF^{-2/3}\right)+{1\over 3}\ell^3+\ell\left(G^{1/3}v+(s+p)G^{-2/3}\right)\right]\right\}.\nonumber\\\label{C3}
\end{eqnarray}
First integrating over $v$ in (\ref{C3}), which leads to a factor $2\pi\delta\left(kF^{1/3}-\ell G^{1/3}\right)$, then integrating over $\ell$, and finally making the substitution $k=[G/(F+G)]^{1/3}q$, we obtain
\begin{eqnarray}
&&\int_{-\infty}^\infty dv\thinspace\psi_{s,F}(-v)
\psi_{s+p,G}(v)=(FG)^{-1/6}(F+G)^{-1/3}\nonumber\\&&\qquad\times{1\over2\pi}\int_{-\infty}^\infty dq\thinspace\exp\left\{ i\left[{1\over 3}q^3+q\left({(s+p)F+sG\over (F+G)^{1/3}(FG)^{2/3}}\right)\right]\right\}\label{C4a}\\
&&\qquad=(FG)^{-1/6}(F+G)^{-1/3}{\rm Ai}\left({(s+p)F+sG\over (F+G)^{1/3}(FG)^{2/3}}\right).\label{C4b}
\end{eqnarray}
In rewriting (\ref{C4a}) in the form (\ref{C4b}), we have again utilized the integral representation (\ref{integralrep}) of the Airy function.

The final expression for $k(F,G)$ in (\ref{kdef}) follows from substituting (\ref{C4b}) in (\ref{C1}). Note that $k(F,G)$ vanishes in the limit $p\to 0$, in accordance with the orthonormality property (\ref{orthonorm2}).

\noindent {\bf Acknowledgements}\\
\noindent I thank Hermann Jo\"el Ouandji Boutcheng, Alberto Rosso, and Andrea Zoia for helpful correspondence.


\begin{thebibliography}{}

\bibitem{OB} Ouandji Boutcheng, H.J., Bouetou, T.B., Burkhardt, T.W., Rosso, A., Zoia, A., Kofane, T.C.: Occupation time statistics of the random acceleration model. J. Stat. Mech. 053213, pp. 1-10 (2016)

\bibitem{TWB93} Burkhardt, T.W., Semiflexible polymer in the half plane and statistics of the integral of a Brownian curve. J. Phys. A  26, L1157-L1162 (1993)

\bibitem{TWB97} Burkhardt, T.W., Free energy of a semiflexible polymer in a tube and statistics of a randomly-accelerated particle. J. Phys. A  30, L167-L172 (1997)
\bibitem{BB1} Bicout, D.J., Burkhardt, T.W., Simulation of a semiflexible polymer in a narrow cylindrical pore. J. Phys. A  34, 5745-5750 (2001)

\bibitem{YBG} Yang, Y., Burkhardt, T.W., Gompper, G., Free energy and extension of a semiflexible polymer in cylindrical confining geometries. Phys. Rev. E 76, 011804, pp. 1-7 (2007)

\bibitem{MB1} Majumdar, S.N., Bray, A.J., Spatial Persistence of Fluctuating Interfaces. Phys. Rev. Lett. 86, 3700-3703 (2001)

\bibitem{GB} Golubovic, L., Bruinsma, R., Surface diffusion and fluctuations of growing interfaces. Phys. Rev. Lett. 66, 321-324 (1991)

\bibitem{DT} Das Sarma, S., Tamborenea, P., A new universality class for kinetic growth: One-dimensional molecular-beam epitaxy. Phys. Rev. Lett. 66, 325-328 (1991)

\bibitem{PV} Valageas, P., Statistical properties of the Burgers equation with Brownian initial velocity. J. Stat. Phys. 134, 589-640 (2009)

\bibitem{McK} McKean, H.P., A winding problem for a resonator driven by a white noise. J. Math. Kyoto Univ. 2, 227-235 (1963) 

\bibitem{MW} Marshall, T.W., Watson, E.J.,  A drop of ink falls from my pen...It comes to earth, I know not when. J. Phys. A 18, 3531-3559 (1985)

\bibitem{Sinai} Sinai, Y.G., Distribution of some functionals of the integral of a random walk. Theor. Math. Phys. 90, 219-241 (1992)

\bibitem{Lachal1} Lachal, A., Les temps de passage successifs  de l'int\'egrale du mouvement brownien. Ann. Inst. Henri Poincar\'e 33, 1-36 (1997)

\bibitem{Lachal2} Lachal, A., Last passage time for integrated Brownian motion. Stoch. Proc. Appl. 49, 57-64 (1994)

\bibitem{DGL} De Smedt, G., Godreche, C., Luck, J.M., Partial survival and inelastic collapse for a randomly accelerated particle. Europhys. Lett. 53, 438-443 (2001)

\bibitem{TWB00} Burkhardt, T.W., Dynamics of Absorption of a Randomly Accelerated Particle. J. Phys. A 33, L429-432 (2000)

\bibitem{FR} Franklin, J. N., Rodemich, E. R., Numerical analysis of an elliptic-parabolic partial differential equation. SIAM J. Numer. Anal. 4, 680-716 (1968)

\bibitem{MP} Masoliver, J., Porr\`a, J. M., Exact solution to the mean exit time problem for free inertial processes driven by Gaussian white noise. Phys. Rev. Lett. 75, 189-192. (1995)

\bibitem{BB2} Bicout, D.J., Burkhardt, T.W., Absorption of a randomly accelerated particle: gambler's ruin in a different game. J. Phys. A  33, 6835-6841 (2000)

 \bibitem{BMS} Bray, A.J., Majumdar, S.N., Schehr, G., Persistence and first-passage properties in nonequilibrium systems. Advances in Physics 62, 225-361 (2013)

\bibitem{TWB14} Burkhardt, T. W.: First Passage of a Randomly Accelerated Particle. In:  Metzler, R., Oshanin, G., Redner, S. (eds.) First-Passage Phenomena and Their Applications, pp. 21-44. World Scientific, Singapore (2014)

\bibitem{Levy} L\'evy, P., Sur certains processus stochastiques homog\`enes. Comp. Math. 7, 283-339 (1939) 

\bibitem{Kac} Kac, M., On distributions of certain Wiener functionals. Trans. Am. Math. Soc. 65, 1-13 (1949) 

\bibitem{Lamperti} Lamperti, J., An occupation time theorem for a class of stochastic processes. Trans. Am. Math. Soc. 88, 380-387 (1958)

\bibitem{Cox} Cox J.T., Griffeath, D., Large deviations for some infinite particle system occupation times. Contemp. Math. 41, 43-54 (1985)

\bibitem{Godreche} Godr\`eche. C., Luck, J.M., Statistics of the occupation time for a random walk in the presence of a moving boundary. 
J. Phys. A 34, 7153-7161 (2001)

\bibitem{Feller} Feller, W., An Introduction to Probability Theory and its Applications. Wiley, New York (1970) 

\bibitem{MRZ} Majumdar, S.N., Rosso, A., Zoia, A., Time at which the maximum of a random acceleration process is reached. J. Phys. A, 43, 115001, pp. 1-16 (2010)

\bibitem{abramowitz}  Abramowitz, M. and Stegun, I. A. (eds.), Handbook of Mathematical Functions. Dover, New York (1965)

\bibitem{risken} Risken, H., The Fokker-Planck Equation: Methods of Solution and Applications, 2nd ed. Springer, Berlin (1989)


\end{thebibliography}
\end{document}